\documentstyle[twoside]{article}

\catcode`\@=11
\long\def\@makefntext#1{
\protect\noindent \hbox to 3.2pt {\hskip-.9pt
$^{{\eightrm\@thefnmark}}$\hfil}#1\hfill}               

\def\@makefnmark{\hbox to 0pt{$^{\@thefnmark}$\hss}}    

\def\ps@myheadings{\let\@mkboth\@gobbletwo
\def\@oddhead{\hbox{}
\rightmark\hfil\eightrm\thepage}
\def\@oddfoot{}\def\@evenhead{\eightrm\thepage\hfil
\leftmark\hbox{}}\def\@evenfoot{}
\def\sectionmark##1{}\def\subsectionmark##1{}}

\oddsidemargin=\evensidemargin
\newcounter{sectionc}\newcounter{subsectionc}\newcounter{subsubsectionc}
\renewcommand{\section}[1] {\vspace{12pt}\addtocounter{sectionc}{1}
\setcounter{subsectionc}{0}\setcounter{subsubsectionc}{0}\noindent
        {\tenbf\thesectionc. #1}\par\vspace{5pt}}
\renewcommand{\subsection}[1] {\vspace{12pt}\addtocounter{subsectionc}{1}
      \setcounter{subsubsectionc}{0}\noindent
      {\bf\thesectionc.\thesubsectionc.{\kern1pt \bfit #1}}\par\vspace{5pt}}
\renewcommand{\subsubsection}[1]
      {\vspace{12pt}\addtocounter{subsubsectionc}{1}
      \noindent{\tenrm\thesectionc.\thesubsectionc.\thesubsubsectionc.
      {\kern1pt \tenit #1}}\par\vspace{5pt}}
\newcommand{\nonumsection}[1] {\vspace{12pt}\noindent{\tenbf #1}
        \par\vspace{5pt}}

\newcounter{appendixc}
\newcounter{subappendixc}[appendixc]
\newcounter{subsubappendixc}[subappendixc]
\renewcommand{\thesubappendixc}{\Alph{appendixc}.\arabic{subappendixc}}
\renewcommand{\thesubsubappendixc}
        {\Alph{appendixc}.\arabic{subappendixc}.\arabic{subsubappendixc}}

\renewcommand{\appendix}[1] {\vspace{12pt}
        \refstepcounter{appendixc}
        \setcounter{figure}{0}
        \setcounter{table}{0}
        \setcounter{lemma}{0}
        \setcounter{theorem}{0}
        \setcounter{corollary}{0}
        \setcounter{definition}{0}
        \setcounter{equation}{0}
        \renewcommand{\thefigure}{\Alph{appendixc}.\arabic{figure}}
        \renewcommand{\thetable}{\Alph{appendixc}.\arabic{table}}
        \renewcommand{\theappendixc}{\Alph{appendixc}}
        \renewcommand{\thelemma}{\Alph{appendixc}.\arabic{lemma}}
        \renewcommand{\thetheorem}{\Alph{appendixc}.\arabic{theorem}}
        \renewcommand{\thedefinition}{\Alph{appendixc}.\arabic{definition}}
        \renewcommand{\thecorollary}{\Alph{appendixc}.\arabic{corollary}}
        \renewcommand{\theequation}{\Alph{appendixc}.\arabic{equation}}
        \noindent{\tenbf Appendix \theappendixc #1}\par\vspace{5pt}}
\newcommand{\subappendix}[1] {\vspace{12pt}
        \refstepcounter{subappendixc}
        \noindent{\bf Appendix \thesubappendixc. {\kern1pt \bfit #1}}
        \par\vspace{5pt}}
\newcommand{\subsubappendix}[1] {\vspace{12pt}
        \refstepcounter{subsubappendixc}
        \noindent{\rm Appendix \thesubsubappendixc. {\kern1pt \tenit #1}}
        \par\vspace{5pt}}

\newcommand{\smalllineskip}{\baselineskip=10pt}

\def\eightcirc{
\begin{picture}(0,0)
\put(4.4,1.8){\circle{6.5}}
\end{picture}}
\def\eightcopyright{\eightcirc\kern2.7pt\hbox{\eightrm c}}

\def\abstracts#1#2#3{{
        \centering{\begin{minipage}{4.5in}\baselineskip=10pt\footnotesize
        \parindent=0pt #1\par
        \parindent=15pt #2\par
        \parindent=15pt #3
        \end{minipage}}\par}}

\renewenvironment{thebibliography}[1]
        {\frenchspacing
         \ninerm\baselineskip=11pt
         \begin{list}{\arabic{enumi}.}
        {\usecounter{enumi}\setlength{\parsep}{0pt}
         \setlength{\leftmargin 12.7pt}{\rightmargin 0pt} 
         \setlength{\itemsep}{0pt} \settowidth
        {\labelwidth}{#1.}\sloppy}}{\end{list}}

\newcounter{itemlistc}
\newcounter{romanlistc}
\newcounter{alphlistc}
\newcounter{arabiclistc}

\newcommand{\fcaption}[1]{
        \refstepcounter{figure}
        \setbox\@tempboxa = \hbox{\footnotesize Fig.~\thefigure. #1}
        \ifdim \wd\@tempboxa > 5in
           {\begin{center}
        \parbox{5in}{\footnotesize\smalllineskip Fig.~\thefigure. #1}
            \end{center}}
        \else
             {\begin{center}
             {\footnotesize Fig.~\thefigure. #1}
              \end{center}}
        \fi}

\newcommand{\tcaption}[1]{
        \refstepcounter{table}
        \setbox\@tempboxa = \hbox{\footnotesize Table~\thetable. #1}
        \ifdim \wd\@tempboxa > 5in
           {\begin{center}
        \parbox{5in}{\footnotesize\smalllineskip Table~\thetable. #1}
            \end{center}}
        \else
             {\begin{center}
             {\footnotesize Table~\thetable. #1}
              \end{center}}
        \fi}

\def\@citex[#1]#2{\if@filesw\immediate\write\@auxout
        {\string\citation{#2}}\fi
\def\@citea{}\@cite{\@for\@citeb:=#2\do
        {\@citea\def\@citea{,}\@ifundefined
        {b@\@citeb}{{\bf ?}\@warning
        {Citation `\@citeb' on page \thepage \space undefined}}
        {\csname b@\@citeb\endcsname}}}{#1}}

\newif\if@cghi
\def\cite{\@cghitrue\@ifnextchar [{\@tempswatrue
        \@citex}{\@tempswafalse\@citex[]}}
\def\citelow{\@cghifalse\@ifnextchar [{\@tempswatrue
        \@citex}{\@tempswafalse\@citex[]}}
\def\@cite#1#2{{$\null^{#1}$\if@tempswa\typeout
        {IJCGA warning: optional citation argument
        ignored: `#2'} \fi}}

\def\@refcitex[#1]#2{\if@filesw\immediate\write\@auxout
        {\string\citation{#2}}\fi
\def\@citea{}\@refcite{\@for\@citeb:=#2\do
        {\@citea\def\@citea{, }\@ifundefined
        {b@\@citeb}{{\bf ?}\@warning
        {Citation `\@citeb' on page \thepage \space undefined}}
        \hbox{\csname b@\@citeb\endcsname}}}{#1}}

\def\@refcite#1#2{{#1\if@tempswa\typeout
        {IJCGA warning: optional citation argument
        ignored: `#2'} \fi}}

\def\refcite{\@ifnextchar[{\@tempswatrue
        \@refcitex}{\@tempswafalse\@refcitex[]}}

\def\pmb#1{\setbox0=\hbox{#1}
        \kern-.025em\copy0\kern-\wd0
        \kern.05em\copy0\kern-\wd0
        \kern-.025em\raise.0433em\box0}

\def\fnt#1#2{\footnotetext{\kern-.3em
        {$^{\mbox{\scriptsize #1}}$}{#2}}}

\def\fpage#1{\begingroup
\voffset=.3in
\thispagestyle{empty}\begin{table}[b]\centerline{\footnotesize #1}
        \end{table}\endgroup}

\def\runninghead#1#2{\pagestyle{myheadings}
\markboth{{\protect\footnotesize\it{\quad #1}}\hfill}
{\hfill{\protect\footnotesize\it{#2\quad}}}}
\font\tenrm=cmr10
\font\tenit=cmti10
\font\tenbf=cmbx10
\font\bfit=cmbxti10 at 10pt
\font\ninerm=cmr9

\font\eightrm=cmr8

\def\qed{\hbox{${\vcenter{\vbox{                      
   \hrule height 0.4pt\hbox{\vrule width 0.4pt height 6pt
   \kern5pt\vrule width 0.4pt}\hrule height 0.4pt}}}$}}

\begin{document}


\runninghead{V. Dvoeglazov}
{Longitudinal Nature $\ldots$}

\thispagestyle{empty}\setcounter{page}{319}
\vspace*{0.88truein}
\fpage{319}

\centerline{\bf `LONGITUDINAL NATURE' OF ANTISYMMETRIC}
\centerline{\bf TENSOR FIELD AFTER QUANTIZATION AND}
\centerline{\bf IMPORTANCE OF THE NORMALIZATION}
\vspace*{0.035truein}

\vspace*{0.37truein}
\centerline{\footnotesize Valeri V. Dvoeglazov}

\centerline{\footnotesize \it
Escuela de F\'{\i}sica, Universidad Aut\'onoma de Zacatecas}
\baselineskip=10pt
\centerline{\footnotesize \it
Apartado Postal C-580, Zacatecas 98068, ZAC., M\'exico}
\baselineskip=10pt
\centerline{\footnotesize \it
E-mail:  valeri@ahobon.reduaz.mx}
\baselineskip=10pt
\centerline{\footnotesize \it URL:
http://ahobon.reduaz.mx/\~~valeri/valeri.htm}

\baselineskip 5mm

\vspace*{0.21truein}

\abstracts{It has long been claimed that the antisymmetric tensor field
of the second rank is pure longitudinal after quantization.
In my opinion, such a situation is quite unacceptable.
I repeat the well-known procedure of the derivation of
the set of Proca equations. It is shown that it can be written
in various forms.   Furthermore, on the basis of the Lagrangian formalism
I calculate dynamical invariants (including the Pauli-Lubanski vector of
relativistic spin for this field).  Even at the classical level
the Pauli-Lubanski vector can be  equal to zero after applications of
well-known constraints.  The importance of the normalization  is pointed
out for the problem of the  description of  quantized fields of
maximal spin 1.  The correct  quantization procedure  permits us to
propose a solution of this puzzle in the modern field theory.  Finally, the
discussion of the connection of the
Ogievetski\u{\i}-Polubarinov-Kalb-Ramond field and  electrodynamic gauge
is presented.}{}{}

\bigskip

$$$$

\section{Introduction}

Quantum electrodynamics (QED) is a construct which has found overwhelming
experimental confirmations (for recent reviews see, {\it e.g.},
refs.~\cite{BS1,BS2}). Nevertheless, a number of theoretical aspects
of this theory deserve more attention. First of all, they are:
the problem of ``fictious photons of helicity other than $\pm j$, as well
as the indefinite metric that must accompany them"; the renormalization
idea, which ``would be sensible only if it was applied with finite
renormalization factors, not infinite ones (one is not allowed to neglect
[and to subtract] infinitely large quantities)"; contradictions with the
Weinberg theorem ``that no symmetric tensor field of rank $j$ can be
constructed from the creation and annihilation operators of massless
particles of spin $j$",\, {\it etc.} They were shown  by
Dirac~\cite{Dirac1,Dirac2} and by Weinberg~\cite{Weinberg}.
Moreover, it appears now that we do not yet understand many specific
features of classical electromagnetism; first of all, the  problems of
longitudinal modes, of the gauge, of the Coulomb action-at-a-distance,
and of the Horwitz' additional invariant parameter,
refs.~\cite{Staru,Evans1,DVO1,DVO2,DVO3,DVO4,DVO5,Chubykalo,Chub,Horwitz}.
Secondly, the standard model, which has been constructed on the basis of
ideas, which are similar to QED, appears to be unable to explain many
puzzles in neutrino physics.

In my opinion, all these shortcomings can be the consequence
of ignoring several important questions.
``In the classical electrodynamics of charged particles, a
knowledge of $F^{\mu\nu}$ completely determines the properties of the
system. A knowledge of $A^\mu$ is redundant there, because it is
determined only up to gauge transformations, which do not affect
$F^{\mu\nu}$\ldots  Such is not the case in quantum
theory\ldots"~\cite{Huang}. We learnt, indeed, about this fact from
the Aharonov-Bohm~\cite{A1} and the Aharonov-Casher effects~\cite{A2}\,.
However, recently several attempts have been undertaken to explain the
Aharonov-Bohm effect classically~\cite{A3}.
These attempts have, in my opinion, logical basis.  In the meantime,
quantizing the antisymmetric tensor field led us to a new puzzle,
which until now had not received fair hearings.  It was claimed that the
antisymmetric tensor field of the second rank is longitudinal after
quantization (in the sense of
the helicity $\sigma=0$),
refs.~\cite{Ogievet,Hayashi,Love,AVD,Sorella}.\,\footnote{M. Kalb and P.
Ramond claimed explicitly~[21b, p. 2283, the third line from below]:
``thus, the massless $\phi_{\mu\nu}$ has one degree of freedom".
While they call $\phi_{\mu\nu}$ as ``potentials" for the field
$F^{\alpha\beta\gamma} = \partial^\alpha \phi^{\beta\gamma}
+\partial^\beta \phi^{\gamma\alpha} +\partial^\gamma \phi^{\alpha\beta}$,
nevertheless, the physical content of the antisymmetric tensor
field of the second rank (the representation $(1,0)\oplus (0,1)$ of the
Lorentz group) must be in accordance with the requirements of relativistic
invariance. Furthermore, ``the helicity -- the projection of the spin onto
the direction of motion -- proves to be equal to zero \ldots even without
the restriction to plane waves, the 3-vector of spin [formula (12)
of~\cite{AVD}] vanishes on solutions \ldots", ref.~[23b], Avdeev and
Chizhov claimed in their turn.}  In the meantime, we
know that the antisymmetric tensor field (electric and magnetic fields,
indeed) is transverse in the Maxwellian classical electrodynamics.
It is not clear how
physically longitudinal components can be transformed into
the physically transverse ones in some
limit. It may be of interest to compare this
question with the group-theoretical consideration in ref.~\cite{Kim} which
deals with the reduction of rotational degrees of freedom to
gauge degrees of freedom in infinite-momentum/zero-mass limit.  See
the only
mentions of the transversality of the quantized antisymmetric tensor field
in refs.~\cite{Takahashi,Boyarkin}. It is often concluded:  one is not
allowed to use the antisymmetric tensor field to represent the quantized
electromagnetic field in relativistic quantum mechanics.
Instead one should pay attention
to the 4-vector potential and gauge freedom. Nevertheless, I am
convinced that a reliable theory should be constructed on the basis of a
minimal number of ingredients (``Occam's Razor") and should have
a well-defined classical limit (as well as massless limit).
Moreover, physicists recently turned again to the problem
of energy in CED~\cite{Chu,Chu-com}. Therefore, in
this paper I undertake a detailed analysis of translational and rotational
properties of the antisymmetric tensor field, I derive {\it various}
forms of the Proca equations (which also can be written in the
Duffin-Kemmer form), then calculate the Pauli-Lubanski operator of
relativistic spin (which must define whether the quantum  is in the left-
or right- polarized states or in the unpolarized state) and  then
conclude, if it is possible to obtain the conventional electromagnetic
theory with photon helicities $\sigma=\pm 1$ provided that strengths ({\it
not} potentials) are chosen to be physical variables in the Lagrangian
formalism.  The particular case also exists when the Pauli-Lubanski vector
for the antisymmetric tensor field of the second rank is equal to zero,
that corresponds to the claimed `{\it longitudinality}' (helicity
$\sigma=0$ ?) of this field. The answer achieved is that the physical
results {\it depend} on the normalization and chosen type of `gauge'
freedom.

Research in this area from a viewpoint of the Weinberg's $2(2j+1)$
component theory have been started in
refs.~\cite{DVO00,DVO01,DVO02,DVO1,DVO2,DVO3,DVO4,DVO5,Gian,Dv-S}.
I would also like to point out that the problem at hand is directly
connected with our understanding of the nature of neutral particles,
including neutrinos~\cite{Majorana,MLC,Ziino,DVA-NP}.
From a mathematical viewpoint, theoretical content provided by the
space-time structure and corresponding symmetries should not depend on
what representation space, which field operators transform on, is chosen.

\section{Bargmann-Wigner Procedure, the Proca Equations
and Relevant Field Functions}

We believe in the power of the group-theoretical methods in the analyses
of the physical behaviour of different-type classical (and quantum)
fields. We also believe that the Dirac equation can be applied to
some particular quantum states of the spin $1/2$. Finally, we believe that
the spin-0 and spin-1 particles can be constructed by taking the direct
product of the spin-1/2 field functions~\cite{BW}. So, on the basis of
these postulates let us firstly repeat the Bargmann-Wigner procedure of
obtaining the equations for bosons of spin 0 and 1.
The set of basic equations for  $j=0$ and $j=1$ are written, e.g.,
ref.~\cite{Lurie}
\begin{eqnarray} \left [
i\gamma^\mu \partial_\mu -m \right ]_{\alpha\beta} \Psi_{\beta\gamma} (x)
&=& 0\quad,\label{bw1}\\ \left [ i\gamma^\mu \partial_\mu -m \right
]_{\gamma\beta} \Psi_{\alpha\beta} (x) &=& 0\quad.\label{bw2}
\end{eqnarray}
We expand the $4\times 4$ matrix wave function into the antisymmetric and
symmetric parts in a standard way
\begin{eqnarray}
\Psi_{[\alpha\beta ]} &=& R_{\alpha\beta} \phi +
\gamma^5_{\alpha\delta} R_{\delta\beta} \widetilde \phi +
\gamma^5_{\alpha\delta} \gamma^\mu_{\delta\tau} R_{\tau\beta} \widetilde
A_\mu \quad,\label{as}\\ \Psi_{\left \{ \alpha\beta \right \}} &=&
\gamma^\mu_{\alpha\delta} R_{\delta\beta} A_\mu
+\sigma^{\mu\nu}_{\alpha\delta} R_{\delta\beta} F_{\mu\nu}\quad,
\label{si}
\end{eqnarray}
where $R= CP$ has the
properties (which are necessary to make expansions (\ref{as},\ref{si}) to
be possible in such a form)
\begin{eqnarray}
&& R^T = -R\quad,\quad R^\dagger =R = R^{-1}\quad,\\
&& R^{-1} \gamma^5 R = (\gamma^5)^T\quad,\\
&& R^{-1} \gamma^\mu R = -(\gamma^\mu)^T\quad,\\
&& R^{-1} \sigma^{\mu\nu} R = - (\sigma^{\mu\nu})^T\quad.
\end{eqnarray}
The  explicit form of this matrix can be chosen:
\begin{equation}
R=\pmatrix{i\Theta & 0\cr 0&-i\Theta\cr}\quad,\quad \Theta = -i\sigma_2 =
\pmatrix{0&-1\cr 1&0\cr},
\end{equation} provided that $\gamma^\mu$
matrices are in the Weyl representation.  The equations
(\ref{bw1},\ref{bw2}) lead to the Kemmer set of the $j=0$ equations:
\begin{eqnarray}
m\phi &=&0 \quad,\\
m\widetilde \phi &=& -i\partial_\mu \widetilde A^\mu\quad,\\
m\widetilde A^\mu &=& -i\partial^\mu \widetilde \phi\quad,
\end{eqnarray}
and to the Proca-Duffin-Kemmer set of the equations for the $j=1$
case:\,\,\footnote{We could use another symmetric matrix $\gamma^5
\sigma^{\mu\nu} R$ in the expansion of the symmetric spinor of the second
rank.  In this case the equations will read
\begin{eqnarray}
&& i\partial_\alpha \widetilde F^{\alpha\mu} +{m\over 2} B^\mu = 0\quad,
\label{de1}\\
&& 2im \widetilde F^{\mu\nu} = \partial^\mu B^\nu -\partial^\nu
B^\mu\quad.\label{de2}
\end{eqnarray}
in which  the dual tensor
$\widetilde F^{\mu\nu}= {1\over 2} \epsilon^{\mu\nu\rho\sigma}
F_{\rho\sigma}$ presents,
because we used that in the Weyl representation
$\gamma^5 \sigma^{\mu\nu} = {i\over 2} \epsilon^{\mu\nu\rho\sigma}
\sigma_{\rho\sigma}$; $B^\mu$ is the corresponding vector potential.  The
equation for the antisymmetric tensor field (which can be obtained from
this set) does not change its form (cf.~\cite{DVO5,DVO96R}) but we see some
``renormalization" of the field functions. In general, it is permitted to
choose various relative phase factors in the expansion of the
symmetric wave function (4) and also consider the matrix term of the form
$\gamma^5 \sigma^{\mu\nu}$.  We shall have additional phase factors in
equations relating the physical fields and the 4-vector potentials.  They
can be absorbed by the redefinition of the potentials/fields (the choice
of normalization/phase).  The above discussion shows that the dual tensor
of the second rank can also be expanded in potentials, as opposed to the
opinion of the referee (JPA) of my previous
paper.}\,\,$^{,}$\,\,\footnote{Recently, after completing this work the
paper~\cite{Kirch} was brought to our attention. It deals with the
redundant components in the $j=3/2$ spin case. If the claims of that paper
are correct we would have to change slightly a verbal terminology which we
use to describe the above equations.} \begin{eqnarray} &&\partial_\alpha
F^{\alpha\mu} + {m\over 2} A^\mu = 0 \quad,\label{1} \\ &&2 m F^{\mu\nu} =
\partial^\mu A^\nu - \partial^\nu A^\mu \quad, \label{2} \end{eqnarray} In
the meantime, in the textbooks, the latter set is usually written as ({\it
e.g.}, ref.~\cite{Itzyk}, p. 135) \begin{eqnarray} &&\partial_\alpha
F^{\alpha\mu} + m^2 A^\mu = 0 \quad, \label{3}\\ && F^{\mu\nu} =
\partial^\mu A^\nu - \partial^\nu A^\mu \quad, \label{4} \end{eqnarray}
The set (\ref{3},\ref{4}) is
obtained from (\ref{1},\ref{2}) after the normalization change $A_\mu
\rightarrow 2m A_\mu$ or $F_{\mu\nu} \rightarrow {1\over 2m} F_{\mu\nu}$.
Of course, one can investigate other sets of equations with different
normalization of the $F_{\mu\nu}$ and $A_\mu$ fields. Are all these
sets of equations equivalent?  As we shall see, to answer this question
is not trivial. The paper~[34a] argued that
the physical normalization is such that in the massless-limit
zero-momentum field functions should vanish in the momentum
representation (there are no massless particles at rest). Next, we
advocate the following approach:  the massless limit can and must be taken
in the end of all calculations only, {\it i.~e.}, for physical quantities.

Let us proceed further. In order to be able to answer the question about
the behaviour of the spin operator
${\bf J}^i = {1\over 2} \epsilon^{ijk}
J^{jk}$ in the massless limit,
one should know the behaviour of the fields $F_{\mu\nu}$ and/or $A_\mu$ in
the massless limit.  We want to analyze the first set (\ref{1},\ref{2}).
If one advocates the following definitions~\cite{Wein} (p. 209)
\begin{eqnarray}
\epsilon^\mu  ({\bf 0}, +1) = - {1\over \sqrt{2}}
\pmatrix{0\cr 1\cr i \cr 0\cr}\,,\quad
\epsilon^\mu  ({\bf 0}, 0) =
\pmatrix{0\cr 0\cr 0 \cr 1\cr}\,,\quad
\epsilon^\mu  ({\bf 0}, -1) = {1\over \sqrt{2}}
\pmatrix{0\cr 1\cr -i \cr 0\cr}\,,
\end{eqnarray}
and ($\widehat p_i = p_i /\mid {\bf p} \mid$,\, $\gamma
= E_p/m$), ref.~\cite{Wein} (p. 68) or ref.~\cite{Novozh} (p. 108),
\begin{eqnarray}
&&
\epsilon^\mu ({\bf p}, \sigma) =
L^{\mu}_{\quad\nu} ({\bf p}) \epsilon^\nu ({\bf 0},\sigma)\,, \\
&& L^0_{\quad 0} ({\bf p}) = \gamma\, ,\quad L^i_{\quad 0} ({\bf p}) =
L^0_{\quad i} ({\bf p}) = \widehat p_i \sqrt{\gamma^2 -1}\, ,\\
&&L^i_{\quad k} ({\bf p}) = \delta_{ik} +(\gamma -1) \widehat p_i \widehat
p_k \quad \end{eqnarray}
for the field operator of the 4-vector
potential, ref.~\cite{Novozh} (p. 109) or
ref.~\cite{Itzyk} \linebreak
(p. 129)\,\footnote{Remember that the invariant integral
measure over the Minkowski space for physical particles is $$\int d^4 p
\delta (p^2 -m^2)\equiv \int {d^3  {\bf p} \over 2E_p}\quad,\quad E_p =
\sqrt{{\bf p}^2 +m^2}\quad.$$ Therefore, we use the field operator as in
(\ref{fo}). The coefficient $(2\pi)^3$ can be considered at this stage as
chosen for convenience.  In ref.~\cite{Wein} the factor $1/(2E_p)$ was
absorbed in creation/annihilation operators and instead of the field
operator (\ref{fo}) the operator was used in which the $\epsilon^\mu
({\bf p}, \sigma)$ functions for a massive spin-1 particle were
substituted by $u^\mu ({\bf p}, \sigma) = (2E_p)^{-1/2} \epsilon^\mu ({\bf
p}, \sigma)$, which may lead to confusion in the definitions of the
massless limit $m\rightarrow 0$ for  classical polarization vectors.}\,
$^,$\,\footnote{In
general, it might be useful to consider front-form helicities (and/or
``time-like" polarizations) too.  But, we leave the presentation of
rigorous theory of this type for subsequent publications.}
\begin{equation} A^\mu (x) =
\sum_{\sigma=0,\pm 1} \int {d^3 {\bf p} \over (2\pi)^3 }
{1\over 2E_p} \left
[\epsilon^\mu ({\bf p}, \sigma) a ({\bf p}, \sigma) e^{-ip\cdot x} +
(\epsilon^\mu ({\bf p}, \sigma))^c b^\dagger ({\bf p}, \sigma) e^{+ip\cdot
x} \right ]\,, \label{fo}
\end{equation}
the normalization of the wave
functions in the momentum representation is thus chosen to the unit,
$\epsilon_\mu^\ast ({\bf p}, \sigma) \epsilon^\mu ({\bf p},\sigma) = -
1$.\footnote{The metric used in this paper $g^{\mu\nu} = \mbox{diag} (1,
-1, -1, -1)$ is different from that of ref.~\cite{Wein}.} \, We observe
that in the massless limit all the defined polarization vectors of the
momentum space do not have good behaviour; the functions describing spin-1
particles tend to infinity.  This is not satisfactory.  Nevertheless,
after renormalizing the potentials, {\it e.~g.}, $\epsilon^\mu \rightarrow
u^\mu \equiv m \epsilon^\mu$ we come to the wave functions in the momentum
representation:\footnote{\,\,It is interesting to note that all the vectors
$u^\mu$ satisfy the condition $p_\mu u^\mu ({\bf p}, \sigma) = 0$.
It is relevant to the case of the Lorentz gauge and, perhaps, to the
analyses of the neutrino theories of light.}
\begin{equation}
u^\mu
({\bf p}, +1)= -{N\over \sqrt{2}m}\pmatrix{p_r\cr m+ {p_1 p_r \over
E_p+m}\cr im +{p_2 p_r \over E_p+m}\cr {p_3 p_r \over
E_p+m}\cr}\,,\quad  u^\mu ({\bf p}, -1)= {N\over
\sqrt{2}m}\pmatrix{p_l\cr m+ {p_1 p_l \over E_p+m}\cr -im +{p_2 p_l \over
E_p+m}\cr {p_3 p_l \over E_p+m}\cr}\,,\label{vp12}
\end{equation}
\begin{equation}
\qquad \qquad \qquad u^\mu ({\bf
p}, 0) = {N\over m}\pmatrix{p_3\cr {p_1 p_3 \over E_p+m}\cr {p_2 p_3
\over E_p+m}\cr m + {p_3^2 \over E_p+m}\cr}\,,  \label{vp3}
\end{equation}
($N=m$ and $p_{r,l} = p_1 \pm ip_2$) which do not
diverge in the massless limit.  Two of the massless functions (with
 $\sigma = \pm 1$) are equal to zero when the particle, described by this
field, is moving along the third axis ($p_1 = p_2 =0$,\, $p_3 \neq 0$).
The third one ($\sigma = 0$) is \begin{equation} u^\mu (p_3, 0)
\mid_{m\rightarrow 0} = \pmatrix{p_3\cr 0\cr 0\cr {p_3^2 \over E_p}\cr}
\equiv  \pmatrix{E_p \cr 0 \cr 0 \cr E_p\cr}\quad, \end{equation}
\setcounter{footnote}{0}
and at
the rest ($E_p=p_3 \rightarrow 0$) also vanishes. Thus, such a field
operator describes the ``longitudinal photons" which is in complete
accordance with the Weinberg theorem $B-A= \sigma$
for massless particles (let us remind  that we
use the $D(1/2,1/2)$ representation). Thus, the change of the
normalization can lead to the ``change" of physical content described by
the classical field (at least, comparing with the well-accepted one).  Of
course, in the quantum case one should somehow fix the form of commutation
relations by some physical principles.\footnote{I am {\it very} grateful
to the anonymous referee of my previous papers (``Foundation of Physics")
who suggested to fix them by requirements of the dimensionless nature of
the action (apart from the requirements of the translational and
rotational invariancies).} \, In the connection with the above
consideration it is interesting to remind that the authors of
ref.~\cite{Itzyk} (see page 136 therein) tried to inforce the
Stueckelberg's Lagrangian in order to overcome the difficulties related to
the $m\rightarrow 0$ limit (or the Proca theory $\rightarrow$ Quantum
Electrodynamics).  The Stueckelberg's Lagrangian is well known to contain
the additional term which may be put in correspondence to some scalar
(longitudinal) field (cf.  also~\cite{Staru}).

Furthermore, it is easy to prove that the physical
fields $F^{\mu\nu}$ (defined as in (\ref{1},\ref{2}),
for instance) vanish in the
massless zero-momentum limit under the both definitions of normalization
and field equations. It is straightforward to find ${\bf
B}^{(+)} ({\bf p}, \sigma ) = {i\over 2m} {\bf p} \times {\bf u}
({\bf p}, \sigma)$, \, ${\bf E}^{(+)} ({\bf p}, \sigma) = {i\over 2m}
p_0 {\bf u} ({\bf p}, \sigma) - {i\over 2m} {\bf p} u^0
({\bf p}, \sigma)$ and the corresponding negative-energy strengths. Here
they are:\footnote{In this paper we assume that $[\epsilon^\mu ({\bf
p},\sigma) ]^c =e^{i\alpha_\sigma} [\epsilon^\mu ({\bf p},\sigma )
]^\ast$, with $\alpha_\sigma$ being arbitrary phase factors at this stage.
Thus, ${\cal C} = I_{4\times 4}$ and $S^c ={\cal K}$.
It is interesting to investigate other choices of the ${\cal C}$,
the charge conjugation matrix and/or consider a field operator composed
of CP-conjugate states.}
\begin{eqnarray}
{\bf B}^{(+)} ({\bf p},
+1) &=& -{iN\over 2\sqrt{2}m} \pmatrix{-ip_3 \cr p_3 \cr ip_r\cr} =
+ e^{-i\alpha_{-1}} {\bf B}^{(-)} ({\bf p}, -1 ) \quad,\quad   \label{bp}\\
{\bf B}^{(+)} ({\bf
p}, 0) &=& {iN \over 2m} \pmatrix{p_2 \cr -p_1 \cr 0\cr} =
- e^{-i\alpha_0} {\bf B}^{(-)} ({\bf p}, 0) \quad,\quad \label{bn}\\
{\bf B}^{(+)} ({\bf p}, -1)
&=& {iN \over 2\sqrt{2} m} \pmatrix{ip_3 \cr p_3 \cr -ip_l\cr} =
+ e^{-i\alpha_{+1}} {\bf B}^{(-)} ({\bf p}, +1)
\quad,\quad\label{bm}
\end{eqnarray}
and
\begin{eqnarray}
{\bf E}^{(+)} ({\bf p}, +1) &=&  -{iN\over 2\sqrt{2}m} \pmatrix{E_p- {p_1
p_r \over E_p +m}\cr iE_p -{p_2 p_r \over E_p+m}\cr -{p_3 p_r \over
E+m}\cr} = + e^{-i\alpha^\prime_{-1}}
{\bf E}^{(-)} ({\bf p}, -1) \quad,\quad\label{ep}\\
{\bf E}^{(+)} ({\bf p}, 0) &=&  {iN\over 2m} \pmatrix{- {p_1 p_3
\over E_p+m}\cr -{p_2 p_3 \over E_p+m}\cr E_p-{p_3^2 \over
E_p+m}\cr} = - e^{-i\alpha^\prime_0} {\bf E}^{(-)} ({\bf p}, 0)
\quad,\quad\label{en}\\
{\bf E}^{(+)} ({\bf p}, -1) &=&  {iN\over
2\sqrt{2}m} \pmatrix{E_p- {p_1 p_l \over E_p+m}\cr -iE_p -{p_2 p_l \over
E_p+m}\cr -{p_3 p_l \over E_p+m}\cr} = + e^{-i\alpha^\prime_{+1}} {\bf
E}^{(-)} ({\bf p}, +1) \quad,\quad\label{em}
\end{eqnarray}
where we denoted, as previously, a normalization factor appearing in the
definitions of the potentials (and/or in the definitions of the physical
fields through potentials) as $N$. Let us note that as a result of the
above definitions we have
\begin{itemize}
\item
The cross products of   magnetic fields of different spin states (such as
${\bf B}^{(+)} ({\bf p}, \sigma) \times {\bf B}^{(-)} ({\bf p},
\sigma^\prime)$) may  be unequal to zero and may be expressed  by
the ``time-like" potential (see the formula (\ref{tp})
below):\footnote{The relevant phase factors are assumed to be equal to zero
here.}
\begin{eqnarray}
{\bf B}^{(+)} ({\bf p}, +1) \times {\bf B}^{(-)} ({\bf p}, +1)
&=& - {iN^2 \over 4m^2} p_3 \pmatrix{p_1\cr p_2\cr
p_3\cr} =\nonumber\\
= -{\bf B}^{(+)} ({\bf p}, -1) \times {\bf B}^{(-)} ({\bf p},
-1)\, ,\\
{\bf B}^{(+)} ({\bf p}, +1) \times {\bf B}^{(-)} ({\bf p}, 0)
&=& - {iN^2 \over 4m^2} {p_r \over \sqrt{2}} \pmatrix{p_1\cr
p_2\cr p_3\cr} =\nonumber\\
= + {\bf B}^{(+)} ({\bf p}, 0) \times {\bf
B}^{(-)} ({\bf p}, -1)\, ,\\
{\bf B}^{(+)} ({\bf p}, -1)
\times {\bf B}^{(-)} ({\bf p}, 0) &=& - {iN^2 \over 4m^2} {p_l \over
\sqrt{2}} \pmatrix{p_1\cr p_2\cr p_3\cr} = \nonumber\\
= + {\bf B}^{(+)} ({\bf p}, 0)
\times {\bf B}^{(-)} ({\bf p}, +1)\,.
\end{eqnarray}
Other
cross products are equal to zero.

\item
Furthermore, one can find the interesting relation:
\begin{eqnarray}
&&{\bf B}^{(+)} ({\bf p}, +1) \cdot {\bf B}^{(-)} ({\bf p},
+1) + {\bf B}^{(+)} ({\bf p}, -1) \cdot {\bf B}^{(-)} ({\bf p}, -1) +
\nonumber\\
&&\qquad\qquad + {\bf
B}^{(+)} ({\bf p}, 0) \cdot {\bf B}^{(-)} ({\bf p}, 0)
={N^2\over 2m^2} (E_p^2 - m^2 )\, ,
\end{eqnarray}
due to
\begin{eqnarray}
&&{\bf B}^{(+)} ({\bf p}, +1) \cdot {\bf B}^{(-)} ({\bf p}, +1) =
{N^2 \over 8m^2} (p_r p_l +2p_3^2) = \nonumber\\
&&\qquad\qquad  = + {\bf B}^{(+)} ({\bf p}, -1)
\cdot {\bf B}^{(-)} ({\bf p}, -1)\, ,\\
&&{\bf B}^{(+)} ({\bf p}, +1) \cdot {\bf B}^{(-)} ({\bf p}, 0) =
{N^2 \over 4\sqrt{2} m^2} p_3 p_r = - {\bf B}^{(+)} ({\bf p}, 0)
\cdot {\bf B}^{(-)} ({\bf p}, -1)\, ,\nonumber\\
&&\\
&&{\bf B}^{(+)} ({\bf p}, -1) \cdot {\bf B}^{(-)} ({\bf p}, 0) =
-{N^2 \over 4\sqrt{2} m^2} p_3 p_l = - {\bf B}^{(+)} ({\bf p}, 0)
\cdot {\bf B}^{(-)} ({\bf p}, +1)\, ,\nonumber\\
&&\\
&&{\bf B}^{(+)} ({\bf p}, +1) \cdot {\bf B}^{(-)} ({\bf p}, -1) =
{N^2 \over 8 m^2} p_r^2\,\, ,\\
&&{\bf B}^{(+)} ({\bf p}, -1) \cdot {\bf B}^{(-)} ({\bf p}, +1) =
{N^2 \over 8 m^2} p_l^2\,\, ,\\
&&{\bf B}^{(+)} ({\bf p}, 0) \cdot {\bf B}^{(-)} ({\bf p}, 0) =
{N^2 \over 4 m^2} p_r p_l\,\, .
\end{eqnarray}

\end{itemize}

For the sake of completeness let us
present the fields corresponding to the ``time-like" polarization:
\begin{equation}
u^\mu ({\bf p}, 0_t) = {N \over m} \pmatrix{E_p\cr p_1
\cr p_2\cr p_3\cr}\quad,\quad {\bf B}^{(\pm)} ({\bf p}, 0_t) = {\bf
0}\quad,\quad {\bf E}^{(\pm)} ({\bf p}, 0_t) = {\bf 0}\,\,.
\label{tp}
\end{equation}
The polarization vector $u^\mu ({\bf p}, 0_t)$ has
the good behaviour in $m\rightarrow 0$, $N=m$ (and also in the subsequent
limit ${\bf p} \rightarrow {\bf 0}$) and it may correspond to some
quantized field (particle).
As one can see, the field operator composed of the states
of longitudinal (e.g., as positive-energy solution) and time-like
(e.g., as negative-energy solution)\footnote{At the present level of our
knowledge only {\it relative} intrinsic parity has physical sense.
Cf.~\cite{DVO5}.}
polarizations may describe a situation when a particle and an antiparticle
have {\it opposite} intrinsic parities (cf.  [34a]).  Furthermore, in the
case of the normalization of potentials to the mass $N=m$  the physical
fields ${\bf B}$ and ${\bf E}$, which correspond to the ``time-like"
polarization, are equal to zero identically.  The longitudinal fields
(strengths) are equal to zero in this limit only when one chooses the
frame  with $p_3 = \mid {\bf p} \mid$, cf. with the light front
formulation, ref.~\cite{DVALF}.  In the case $N=1$ and (\ref{1},\ref{2})
we have, in general, the divergent behaviour of potentials and
strengths.\footnote{In the case of $N=1$ the fields ${\bf B}^\pm ({\bf p},
0_t)$ and ${\bf E}^\pm ({\bf p}, 0_t)$ would be undefined. This fact was
also not fully appreciated in the previous formulations of the theory of
$(1,0)\oplus(0,1)$ and $(1/2,1/2)$ fields.}

\section{Translational and Rotational Properties of Antisymmetric
Tensor Field.}

I begin this Section with the antisymmetric tensor field operator (in
general, complex-valued):
\begin{equation}
F^{\mu\nu} (x) \,=\, \sum_{\sigma=0,\pm 1}
\int \frac{d^3 {\bf p}}{(2\pi)^3 2E_p} \, \left [
F^{\mu\nu}_{(+)} ({\bf p},\sigma)\, a ({\bf p},\sigma)\, e^{-ip x} +
F^{\mu\nu}_{(-)} ({\bf p},\sigma) \,b^\dagger ({\bf p},\sigma)\,
e^{+ip x} \right ]\label{fop}
\end{equation}
and with the Lagrangian,
including, in general, mass term:\footnote{The massless limit
($m\rightarrow 0$) of the Lagrangian is connected with the Lagrangians
used in the conformal field theory and in the conformal supergravity by
adding the total derivative:
\begin{equation} {\cal L}_{CFT} = {\cal L} +
{1\over 2}\partial_\mu \left ( F_{\nu\alpha} \partial^\nu F^{\mu\alpha} -
F^{\mu\alpha} \partial^\nu F_{\nu\alpha} \right )\quad.  \end{equation}
The Kalb-Ramond gauge-invariant
form (with
respect to ``gauge" transformations $F_{\mu\nu}  \rightarrow F_{\mu\nu}
+\partial_\nu \Lambda_\mu - \partial_\mu \Lambda_\nu$),
ref.~\cite{Ogievet,Hayashi}, is obtained only if one uses the Fermi
procedure {\it mutatis mutandis} by removing the additional ``phase" field
$\lambda (\partial_\mu F^{\mu\nu})^2$, with the appropriate coefficient
$\lambda$, from the Lagrangian. This has certain analogy with the QED,
where the question of whether the Lagrangian is gauge-invariant or not, is
solved depending on the presence of the term $\lambda (\partial_\mu
A^\mu)^2$. For details see ref.~\cite{Hayashi} and what is below.

In general it is possible to introduce various forms of the mass term
and of corresponding normalization of the field.}
\begin{equation}\label{Lagran}
{\cal L} =  {1\over 4} (\partial_\mu F_{\nu\alpha})(\partial^\mu
F^{\nu\alpha}) - {1\over 2} (\partial_\mu F^{\mu\alpha})(\partial^\nu
F_{\nu\alpha}) - {1\over 2} (\partial_\mu F_{\nu\alpha})(\partial^\nu
F^{\mu\alpha}) + {1\over 4} m^2 F_{\mu\nu} F^{\mu\nu} \,.
\end{equation}
The Lagrangian leads to the equation of motion in the
following form (provided that the appropriate antisymmetrization
procedure has been taken into account):
\begin{equation} {1\over 2} ({\,\lower0.9pt\vbox{\hrule
\hbox{\vrule height 0.2 cm \hskip 0.2 cm \vrule height
0.2cm}\hrule}\,}+m^2) F_{\mu\nu} +
(\partial_{\mu}F_{\alpha\nu}^{\quad,\alpha} -
\partial_{\nu}F_{\alpha\mu}^{\quad,\alpha}) = 0 \quad,\label{PE}
\end{equation}
where ${\,\lower0.9pt\vbox{\hrule \hbox{\vrule height 0.2 cm
\hskip 0.2 cm
\vrule height 0.2 cm}\hrule}\,}
=- \partial_{\alpha}\partial^{\alpha}$, cf. with the set of equations
(15,16).  It is this equation for antisymmetric-tensor-field components
that follows from the Proca-Duffin-Kemmer-Bargmann-Wigner
consideration,
provided that $m\neq 0$ and in the final expression one takes into account
the Klein-Gordon equation $({\,\lower0.9pt\vbox{\hrule \hbox{\vrule height
0.2 cm \hskip 0.2 cm \vrule height 0.2 cm}\hrule}\,} - m^2) F_{\mu\nu}=
0$.  The latter expresses relativistic dispersion relations $E^2 -{\bf p}^2
=m^2$ and it follows from the coordinate Lorentz transformation
laws~\cite{Ryder}, \S 2.3.

Following the variation procedure given, {\it e.g.}, in
refs.~\cite{Corson,Barut,Bogoliubov} one can obtain that
the energy-momentum tensor is expressed:
\begin{eqnarray}
\Theta^{\lambda\beta} &=& {1\over 2} \left [
(\partial^\lambda F_{\mu\alpha}) (\partial^\beta F^{\mu\alpha})
- 2(\partial_\mu F^{\mu\alpha}) (\partial^\beta F^\lambda_{\quad\alpha}) -
\right.\nonumber\\
&-& \left . 2 (\partial^\mu F^{\lambda\alpha}) (\partial^\beta
F_{\mu\alpha})\right ] -{\cal L} g^{\lambda\beta}\, .
\end{eqnarray}
One can also obtain that
for rotations $x^{\mu^\prime} = x^\mu + \omega^{\mu\nu} x_\nu$
the corresponding variation of the wave function is found
from the formula:
\begin{equation}
\delta F^{\alpha\beta} = {1\over 2} \omega^{\kappa\tau}
{\cal T}_{\kappa\tau}^{\alpha\beta,\mu\nu} F_{\mu\nu}\quad.
\end{equation}
The generators of infinitesimal transformations are then defined as
\begin{eqnarray}
\lefteqn{{\cal T}_{\kappa\tau}^{\alpha\beta,\mu\nu} \,=\,
{1\over 2} g^{\alpha\mu} (\delta_\kappa^\beta \,\delta_\tau^\nu \,-\,
\delta_\tau^\beta\,\delta_\kappa^\nu) \,+\,{1\over 2} g^{\beta\mu}
(\delta_\kappa^\nu\delta_\tau^\alpha  \,-\,
\delta_\tau^\nu\, \delta_\kappa^\alpha) +\nonumber}\\
&+&\,
{1\over 2} g^{\alpha\nu} (\delta_\kappa^\mu \, \delta_\tau^\beta \,-\,
\delta_\tau^\mu \,\delta_\kappa^\beta) \,+\, {1\over 2}
g^{\beta\nu} (\delta_\kappa^\alpha \,\delta_\tau^\mu \,-\,
\delta_\tau^\alpha \, \delta_\kappa^\mu)\quad.
\end{eqnarray}
It is ${\cal T}_{\kappa\tau}^{\alpha\beta,\mu\nu}$, the generators of
infinitesimal transformations,
that enter in the formula for the relativistic spin tensor:
\begin{equation}
J_{\kappa\tau} = \int d^3 {\bf x} \left [ \frac{\partial {\cal
L}}{\partial ( \partial F^{\alpha\beta}/\partial t )} {\cal
T}^{\alpha\beta,\mu\nu}_{\kappa\tau} F_{\mu\nu} \right ]\quad.
\label{inv}
\end{equation}
As a result one  obtains:
\begin{eqnarray}
J_{\kappa\tau} &=& \int d^3 {\bf x} \left [ (\partial_\mu F^{\mu\nu})
(g_{0\kappa} F_{\nu\tau} - g_{0\tau} F_{\nu\kappa}) -  (\partial_\mu
F^\mu_{\,\,\,\,\kappa}) F_{0\tau} + (\partial_\mu F^\mu_{\,\,\,\,\tau})
F_{0\kappa} + \right. \nonumber\\
&+& \left. F^\mu_{\,\,\,\,\kappa} ( \partial_0 F_{\tau\mu} +
\partial_\mu F_{0\tau} +\partial_\tau F_{\mu 0})  -   F^\mu_{\,\,\,\,\tau}
( \partial_0 F_{\kappa\mu} +\partial_\mu F_{0\kappa} +\partial_\kappa
F_{\mu 0}) \right ]\,. \label{gene10}
\end{eqnarray}
If one agrees that the
orbital part of the angular momentum
\begin{equation} L_{\kappa\tau} =
x_\kappa \Theta_{0\,\tau} - x_\tau \Theta_{0\,\kappa} \quad,
\end{equation}
with  $\Theta_{\tau\lambda}$ being the energy-momentum tensor, does not
contribute to the Pauli-Lubanski operator when acting on the
one-particle free states (as in the Dirac $j=1/2$ case), then
the Pauli-Lubanski 4-vector is constructed as
follows~\cite{Itzyk} (Eq. (2-21))
\begin{equation}
W_\mu = -{1\over 2}  \epsilon_{\mu\kappa\tau\nu} J^{\kappa\tau} P^\nu \quad,
\end{equation}
with $J^{\kappa\tau}$ defined by Eqs.
(\ref{inv},\ref{gene10}). The 4-momentum operator $P^\nu$ can be replaced
by its eigenvalue when acting on the plane-wave eigenstates.
Furthermore, one should
choose space-like normalized vector $n^\mu n_\mu = -1$, for example $n_0
=0$,\, ${\bf n} = \widehat  {\bf p} = {\bf p} /\vert {\bf
p}\vert$.\,\,\footnote{\,\, One should remember that the helicity operator
is usually connected with the Pauli-Lubanski vector in the following
manner $({\bf J} \cdot \widehat {\bf p}) = ({\bf W} \cdot \widehat {\bf
p})/ E_p$, see ref.~\cite{Shirok}. The choice of ref.~\cite{Itzyk}, p.
147, $n^\mu = \left ( t^\mu - p^\mu {p\cdot t \over m^2} \right ) {m\over
\mid {\bf p} \mid}$, with $t^\mu \equiv (1,0,0,0)$ being a time-like
vector, is also possible but it leads to some obscurities in the procedure
of taking the massless limit. These obscurities will be clarified in a
separate paper.} \,\, After lengthy calculations in a spirit
of~\cite{Itzyk}, pp.  58, 147, one can find the explicit form of the
relativistic spin:  \begin{eqnarray} && (W_\mu \cdot n^\mu) = - ({\bf
W}\cdot {\bf n}) = -{1\over 2} \epsilon^{ijk} n^k J^{ij}
p^0\quad,\label{PL1}\\
&& {\bf J}^k = {1\over 2} \epsilon^{ijk} J^{ij} =
\epsilon^{ijk} \int d^3 {\bf x} \left [ F^{0i} (\partial_\mu F^{\mu j}) +
F_\mu^{\,\,\,\,j} (\partial^0 F^{\mu i} +\partial^\mu F^{i0} +\partial^i
F^{0\mu} ) \right ]\,.\nonumber\\
&&\label{PL2}
\end{eqnarray}
Now it becomes obvious that the application of the generalized Lorentz
conditions (which are quantum versions of free-space dual Maxwell's
equations) leads in such a formulation to the absence of electromagnetism
in a conventional sense.  The resulting Kalb-Ramond field is longitudinal
(helicity $\sigma=0$).  All the components of the angular momentum tensor
for this case are identically equated to zero. The discussion of this fact
can also be found in ref.~\cite{Hayashi,DVO2}. This situation can occur in
the particular choice of the normalization of the field operators
and unusual ``gauge" invariance.

Furthermore, the spin operator recasts in the terms of the vector
potentials as follows (if one takes into account the dynamical equations,
Eqs.  (\ref{de1},\ref{de2},\ref{1},\ref{2}))\footnote{\,\, The formula
(\ref{spinf}) has certain similarities with the formula for the spin
vector obtained from Eqs.  (5.15,5.21) of ref.~\cite{Bogoliubov}:
\begin{eqnarray}
{\bf J}_i &=& \epsilon_{ijk} \int J_{jk}^0 d^3 {\bf
x}\quad,\\ J_{\alpha\beta}^0 &=& \left ( A_\beta {\partial A_\alpha \over
\partial x_0} - A_\alpha {\partial A_\beta \over \partial x_0} \right
)\,.
\end{eqnarray}
It describes the ``transverse photons" in the ordinary wisdom.
But, not all the questions related with the second $B_\mu$ potential,
the dual tensor $\widetilde{F}^{\mu\nu}$
and the normalization of 4-potentials and fields
have been clarified in the standard textbooks.}
\begin{eqnarray}
{\bf J}^k &=&  \epsilon^{ijk} \int d^3 {\bf x} \left [
F^{0i} (\partial_\mu F^{\mu j} ) + \widetilde F^{0i} (\partial_\mu
\widetilde F^{\mu j}) \right ] = \nonumber\\
&\qquad&\qquad = {1\over 4} \epsilon^{ijk} \int d^3 x \left [ B^j (
\partial^0 B^i - \partial^i B^0 ) - A^j ( \partial^0 A^i -\partial^i A^0 )
\right ] \quad.  \label{spinf}
\end{eqnarray}
If we put, as usual,
$\widetilde F^{\mu\nu} = \pm i F^{\mu\nu}$ (or $B^\mu =\pm A^\mu$) for the
right-  and left- circularly polarized radiation we shall again
obtain equating the spin operator to zero. The same situation would
occur if one chooses ``unappropriate" normalization and/or if one uses the
equations (\ref{3},\ref{4}) in the massless limit
without necessary precautions.  The
straightforward application of (\ref{3},\ref{4}) would lead
to the proportionality $J_{\kappa\tau} \sim m^2$ and, thus, it
appears that the spin operator would be equal to zero in the massless
limit, provided that the components of $A_\mu$ have good behaviour (do not
diverge in $m\rightarrow 0$).  Probably, this fact (the relation between
generators and the normalization)  was the origin of why many respectable
persons claimed that the antisymmetric tensor field is a pure
longitudinal field. On the other hand, in a private communication Prof.
Y.  S.  Kim stressed that neither he nor E. Wigner used the normalization
of the spin generators to the mass.  What is the situation which is
realized in Nature (or both)? The theoretical answer  depends on the
choice of the field operators, namely on the choice of positive- and
negative- energy solutions, creation/annihilation  operators and the
normalization.

One of the possible ways to  obtain helicities $\sigma=\pm 1$
is a modification of the electromagnetic field tensor,
{\it i.e.}, introducing the non-Abelian
electrodynamics~\cite{Evans1,EV-DVA}:
\begin{equation} F_{\mu\nu}\quad
\Rightarrow \quad {\bf G}_{\mu\nu}^{a} = \partial_\mu A_\nu^{(a)} -
\partial_\nu A_\mu^{(a)} -i{e\over \hbar}[ A_\mu^{(b)},
A_\nu^{(c)}] \quad,
\end{equation}
where $(a),\,(b),\,(c)$ denote the vector components in the
$(1),\,(2),\,(3)$ circular basis. In other words, one can add some
ghost field (the ${\bf B}^{(3)}$ field) to the antisymmetric  tensor
$F_{\mu\nu}$ which initially supposed to contain transverse components
only.  As a matter of fact this induces hypotheses on a massive photon
and/or an additional displacement current.  I can agree with the {\it
possibility} of the ${\bf B}^{(3)}$ field concept (while it is required
{\it rigorous} elaboration in the terminology of the modern quantum field
theory), but, at the moment, I prefer to avoid any auxiliary constructions
(even if they may be valuable in intuitive explanations and
generalizations).  If these non-Abelian constructions exist they should be
deduced from a more general theory on the basis of some fundamental
postulates, {\it e.g.}, in a spirit of refs.~\cite{DV-CP,Dv-S,DVA-NP}.
Moreover,  this concept appears to be in contradiction with the
concept of the $m\rightarrow 0$ group contraction for a photon as
presented by Wigner and Inonu~\cite{WI} and Kim~\cite{Kim}.

In the procedure of the
quantization  one can reveal an important case, when the transversality
(in the meaning of existence of $\sigma=\pm 1$) of the antisymmetric
tensor field is preserved.  This conclusion is related with existence of
the dual tensor $\widetilde F^{\mu\nu}$ or with  correcting the procedure
of taking the massless limit.

\setcounter{footnote}{0}

In this Section, I first choose the field operator, Eq. (\ref{fop}), such
that:
\begin{eqnarray} F^{i0}_{(+)} ({\bf p}) &=& E^i ({\bf
p})\quad,\quad F^{jk}_{(+)} ({\bf p}) = - \epsilon^{jkl} B^l ({\bf
p})\quad;\\ F^{i0}_{(-)} ({\bf p}) &=& \tilde F^{i0} ({\bf p}) = B^i ({\bf
p})\quad,\quad F^{jk}_{(-)} ({\bf p}) = \tilde F^{jk} ({\bf p}) =
\epsilon^{jkl} E^l ({\bf p})\quad,
\end{eqnarray}
where
$\tilde F^{\mu\nu} = {1\over 2} \epsilon^{\mu\nu\rho\sigma}
F_{\rho\sigma}$ is the tensor dual to $F^{\mu\nu}$; and
$\epsilon^{\mu\nu\rho\sigma} = - \epsilon_{\mu\nu\rho\sigma}$\, , \,
$\epsilon^{0123} = 1$ is the totally antisymmetric Levi-Civita tensor.
After lengthy but standard calculations one achieves:\footnote{Of course,
the question of the behaviour of vectors ${\bf E}_\sigma ({\bf p})$ and
${\bf B}_\sigma ({\bf p})$ and/or of creation and annihilation operators
with respect to the discrete symmetry operations in this particular case
deserves detailed elaboration.}
\begin{eqnarray} \lefteqn{{\bf J}^k =
\sum_{\sigma\sigma^\prime}\int \frac{d^3 {\bf p}}{(2\pi)^3 2E_p} \left \{
\frac{i\epsilon^{ijk}{\bf E}^i_\sigma ({ \bf p}) {\bf B}^j_{\sigma^\prime}
({\bf p})}{2} \left [ a ({\bf p},\sigma) b^\dagger ({\bf
p},\sigma^\prime ) + a ({\bf p},\sigma^\prime)
b^\dagger ({\bf p},\sigma) + \right .  \right .\nonumber}\\ &+& \left .
\left .  b^\dagger ({\bf p},\sigma^\prime ) a ({\bf p},\sigma ) +
b^\dagger ({\bf p},\sigma) a ({\bf p},\sigma^\prime ) \right ] -
\right.\nonumber\\ &-&\left. {1\over 2E_p} \left [ i{\bf p}^k ({\bf
E}_\sigma ({\bf p}) \cdot {\bf E}_{\sigma^\prime} ({\bf p}) + {\bf
B}_\sigma ({\bf p}) \cdot {\bf B}_{\sigma^\prime} ({\bf p})) -
\right. \right. \\
&-& \left. \left.  i {\bf
E}^k_{\sigma^\prime} ({\bf p}) ({\bf p}\cdot {\bf E}_{\sigma} ({\bf p})) -
i {\bf B}^k_{\sigma^\prime} ({\bf p}) ({\bf p}\cdot {\bf B}_{\sigma} ({\bf
p}))\right ]\times \left [ a ({\bf
p}, \sigma ) b^\dagger ({\bf p},\sigma^\prime) + b^\dagger ({\bf
p},\sigma ) a ({\bf p},\sigma^\prime )\right ] \right \} \nonumber
\end{eqnarray}
One should choose normalization conditions for field functions in the
momentum representation.  For instance, one can use the analogy with the
(dual) classical electrodynamics:\footnote{Different choices of the
normalization can still lead to equating the spin operator to zero or even
to the other values of helicity, which differ from $\pm 1$. This was also
discussed with Prof. N. Mankoc-Borstnik during the Workshop ``Lorentz
Group, CPT and Neutrinos" (Zacatecas, M\'exico, June 23-26, 1999). The
question is:  which cases are realized in Nature and what processes do
correspond to every case?}
\begin{eqnarray} &&({\bf E}_\sigma ({\bf p})
\cdot {\bf E}_{\sigma^\prime} ({\bf p}) + {\bf B}_\sigma ({\bf p}) \cdot
{\bf B}_{\sigma^\prime} ({\bf p}))  = 2E_p
\delta_{\sigma\sigma^\prime}\quad,\\ &&{\bf E}_\sigma \times {\bf
B}_{\sigma^\prime} = {\bf p}\delta_{\sigma\sigma^\prime} -{\bf
p}\delta_{\sigma, -\sigma^\prime}\quad.  \end{eqnarray} These conditions
still imply that ${\bf E}\perp {\bf B} \perp {\bf p}$.

Finally, one obtains
\begin{equation} {\bf J}^k =  - i\sum_\sigma \int \frac{d^3 {\bf
p}}{(2\pi)^3} \, \frac{{\bf p}^k}{2E_p} \left [ a ({\bf p},\sigma)
b^\dagger ({\bf p},-\sigma ) +b^\dagger ({\bf p},\sigma) a
({\bf p},-\sigma ) \right ]\,.\label{spin10}
\end{equation}
If we want to
describe states with the definite helicity quantum number (photons) we
should assume that $b^\dagger ({\bf p},\sigma)= i a^\dagger ({\bf
p},\sigma)$ which is reminiscent of the Majorana-like
theories~\cite{Majorana,DVA-NP}.\,\footnote{Of course, the imaginary unit
can be absorbed by the corresponding re-definition of negative-frequency
solutions.} \, One can take into account the prescription of the normal
ordering and set up the commutation relations in the form:
\begin{equation}
\left [a ({\bf p},\sigma), a^\dagger ({\bf
k},\sigma^\prime )\right ]_{-} = (2\pi)^3 \delta ({\bf p}-{\bf k})
\delta_{\sigma,-\sigma^\prime}\quad.  \label{cr} \end{equation} After
acting the operator (\ref{spin10}) on the physical states, {\it e.g.},
$a^\dagger ({\bf p},\sigma) \vert 0>$ , we are convinced that
the antisymmetric tensor field can describe particles with helicities to
be equal to $\pm 1$). One can see that an origin of this conclusion is the
possibility of different definitions of the field operator (and its
normalization), non-unique definition of the energy-momentum
tensor~\cite{Barut,Chu,Chu-com} and possible existence of the `{\it
antiparticle}' for the  particle described by antisymmetric tensor field.
This consideration is obviously related to the Weinberg discussion of the
connection between helicity and representations of the Lorentz group~[5a].
Next, I would like to point out that the Proca-like equations for
antisymmetric tensor field with {\it mass}, {\it e.g.}, Eq.  (\ref{PE})
can possess tachyonic solutions, see for the discussion in
ref.~\cite{DVO1}.  Therefore, in a massive case the free physical states
can be mixed with unphysical (at the present level of our
knowledge) tachyonic states.

\bigskip
\bigskip

\section{Normalization and $m\rightarrow 0$ Limit of the Proca Theory}

As opposed to the previous sections, where we assumed
non-tentability of the application of the generalized Lorentz condition,
in this Section we pay more attention to the correct procedure of
taking the massless limit.
We note that not all the obscurities were clarified  in
previous sections and recent
works~\cite{Ohanian,DVO96,DVO96R}.\footnote{First of all, we note that the
equality of the angular momentum generators to zero can be re-interpreted
as $$W_\mu P^\mu =0\,\, ,$$ with $W_\mu$ being the Pauli-Lubanski
operator.  This yields $$W_\mu = \lambda P_\mu\,\, $$
in the massless case. But, according to the analysis above the 4-vector
$W_\mu$ would be equal to zero {\it identically} in the massless limit.
This is not satisfactory from the conceptual viewpoints.} \,\, Let us
analyze in a straightforward manner the operator (\ref{spinf}).  If one
uses the following definitions of positive- and negative-energy parts of
the antisymmetric tensor field in the momentum space, i.~e., according to
(\ref{bp}-\ref{em}) with ($\alpha_\sigma =0$):  \begin{equation}
(F_{\mu\nu})^{(+)}_{+1} = + (F_{\mu\nu})^{(-)}_{-1} \, , \,
(F_{\mu\nu})^{(+)}_{-1} = + (F_{\mu\nu})^{(-)}_{+1} \, , \,
(F_{\mu\nu})^{(+)}_{0} =  - (F_{\mu\nu})^{(-)}_{0} \quad.
\end{equation}
for the field operator (\ref{fop})
then one obtains in the frame where ${\bf p}^{1,2} =0$:
\begin{eqnarray}
{\bf J}^k &\equiv& {m\over 2}\int d^3 {\bf x}\,
{\bf E}(x^\mu) \times {\bf A} (x^\mu) =
{m^2 \over 4} \int {d^3 {\bf p} \over (2\pi)^3 \, 4E_p^2}
\left \{ \pmatrix{0\cr 0\cr E_p\cr} \right . \label{fin} \\
&&\left .\left [ a ({\bf p}, +1) b^\dagger ({\bf p},
+1) - a ({\bf p}, -1) b^\dagger ({\bf p}, -1) + \right .\right .
\nonumber\\
&+&\left.\left. b^\dagger ({\bf p}, +1) a
({\bf p}, +1) - b^\dagger ({\bf p}, -1) a ({\bf p}, -1)\right
] +\right .\nonumber\\
&+& \left. {E_p \over m \sqrt{2}} \pmatrix{E_p\cr iE_p \cr 0\cr}
\left [ a ({\bf p},+1) b^\dagger ({\bf p}, 0) +
b^\dagger ({\bf p}, -1) a ({\bf p}, 0)\right ] +\right .\nonumber\\
&+&\left . {E_p \over m \sqrt{2}} \pmatrix{E_p \cr -iE_p \cr 0\cr}
\left [ a ({\bf p}, -1) b^\dagger ({\bf p}, 0) +
b^\dagger ({\bf p}, +1) a ({\bf p}, 0) \right ] +\right .\nonumber\\
&+&\left . {1\over \sqrt{2}} \pmatrix{m\cr -im\cr 0\cr}
\left [ a ({\bf p}, 0) b^\dagger ({\bf p}, +1) +
b^\dagger ({\bf p}, 0)  a ({\bf p}, -1) \right ] +\right .\nonumber\\
&+&\left . {1\over \sqrt{2}} \pmatrix{m\cr im\cr 0\cr}
\left [ a ({\bf p}, 0) b^\dagger ({\bf p}, -1) +
b^\dagger ({\bf p}, 0) a ({\bf p}, +1) \right ]
\right \}\,\, .\nonumber
\end{eqnarray}
Above, we used that according to dynamical equations (15,16)
written in the momentum representation
\begin{eqnarray}
&& [(\partial_\mu F^{\mu j} ({\bf p},\sigma )]^{(+)}
= -{m\over 2} u^j ({\bf p},\sigma)\,\, ,\,\,
[(\partial_\mu F^{\mu j} ({\bf p},\sigma )]^{(-)}
= -{m\over 2} [u^j ({\bf p},\sigma)]^\ast \\
&& [\partial_\mu \widetilde F^{\mu j} ({\bf p},\sigma) ]^\pm =0\,\, .
\end{eqnarray}
Next, it is obvious that though $\partial_\mu F^{\mu\nu}$
may be equal to zero in the massless limit from the formal viewpoint,
and the equation (\ref{fin}) is proportional to the squared mass (?) at
the first sight, it must not be forgotten that the commutation
relations may provide additional mass factors in the denominator of
(\ref{fin}).  It is the factor $\sim E_p/m^{2}$ in the commutation
relations\footnote{Remember that the dimension of the $\delta$ function is
inverse to its argument.}
\begin{equation} [a ({\bf p},\sigma ) ,
b^\dagger ({\bf k},\sigma^\prime ) ] \sim (2\pi)^3 {E_p\over m^2}
\delta_{\sigma\sigma^\prime} \delta ({\bf p} - {\bf k}) \,\, .
\end{equation}
which is required by the
principles of the rotational and translational
invariance\,\footnote{That is to say: the factor $\sim {1\over m^2}$
is required if one wants to obtain non-zero energy (and, hence,
helicity) excitations.} (and
also by the necessity of the description of the Coulomb long-range force
$\sim 1/r^2$ by means of the antisymmetric tensor field of the second
rank).

\setcounter{footnote}{0}

The dimension of the creation/annihilation operators of the 4-vector
potential should be [energy]$^{-2}$ provided that we use
(\ref{vp12},\ref{vp3}) with $N=m$ and $\epsilon^\mu \rightarrow
u^\mu$.  Next, if we want the $F^{\mu\nu}
(x^\mu)$ to have the dimension [energy]$^2$ in the unit system $c=\hbar
=1$,\footnote{The dimensions [energy]$^{+1}$ of the field operators for
strengths was used in my previous papers in order  to keep
similarities with the Dirac case (the $(1/2,0)\oplus (0,1/2)$
representation) where the mass term presents explicitly in the term of the
bilinear combination of the fields.}
\setcounter{footnote}{0}
we must divide the
Lagrangian by $m^2$ (with the same $m$, the mass of
the particle!):
\begin{equation}
{\cal L} = \frac{{\cal L} (Eq. 46)}{m^2}\, .
\end{equation}
In this case, the antisymmetric tensor field has the dimension
which is compatible with the inverse-square law, but the procedure of
taking massless limit is somewhat different (and cannot be
carried out from the beginning).
This procedure will have the influence on the form of
(\ref{spinf},\ref{fin}) because the derivatives in this case pick up the
additional mass factor.  Thus, we can remove the ``ghost" proportionality
of the $c$- number coefficients in (\ref{fin}) to $\sim m$. The
commutation relations also change their form.
For instance, one can now consider that
$[a ({\bf p},\sigma), b^\dagger ({\bf k}, \sigma^\prime )]_-
\sim (2\pi)^3 2E_p \delta_{\sigma\sigma^\prime} \delta ({\bf p}
-{\bf k})$\, .
The possibility of the above
renormalizations was {\it not} noted in the previous papers on the theory
of the 4-vector potential and  of the antisymmetric tensor field of the
second rank. Probably, this  was the reason  why people were confused
after including the mass factor of the creation/annihilation operators in
the field functions of $(1/2,1/2)$ and/or $(1,0)\oplus (0,1)$
representations, and/or applying the generalized Lorentz condition
inside the dynamical invariant(s),
which, as noted above, coincides in the form with the Maxwell free-space
equations.

Finally, we showed that the interplay between definitions of
field functions, Lagrangian and commutation relations occurs, thus
giving the {\it non-zero} values of the angular momentum generators in the
$(1,0)\oplus (0,1)$ representation.

The conclusion of the ``transversality" (in the meaning of existence of
$\sigma=\pm 1$) is in accordance with the conclusion of the Ohanian's
paper~\cite{Ohanian}, cf.  formula (7) there:\footnote{The formula (7) of
ref.~\cite{Ohanian} is in the SI unit system and our arguments above are
similar in the physical content. But, remember that in
almost all papers the electric field is defined to be equal to ${\bf E}^i
= F^{i0} = \partial^i A^0 - \partial^0 A^i$, with the potentials being not
well-defined in the massless limit of the Proca theory.  Usually, the
divergent part of the potentials was referred to the gauge-dependent part.
Furthermore, the physical fields and potentials were considered
classically in the cited paper, so the integration over the 3-momenta (the
quantization inside a cube) was not implied, see the formula (5) there.
Please pay also attention to the complex conjugation operation on
the potentials in the Ohanian's formula.
We did not still exclude the possibility
of the mathematical framework, which is different from our
presentation, but the conclusions, in my opinion,
must be in accordance with the Weinberg theorem.}
\begin{equation} {\bf J} = {1\over 2\mu_0 c^2} \int \Re e ({\bf
E} \times {\bf A}^\ast ) \, d^3 {\bf x} = \pm {1\over \mu_0 c^2} \int
\frac{\hat {\bf z} E_0^2}{\omega} d^3 {\bf x} \,\, ,\label{spin}
\end{equation} with the Weinberg theorem, also with known experiments.
The question, whether the situation could be realized
when the spin of the antisymmetric tensor field would be equal to zero
(in other words, whether the antisymmetric tensor field with
unusual normalization exists
or whether the third state of the massless 4-vector potential exists,
as argued by Ogievetski\u{\i} and Polubarinov~\cite{Ogievet}),
must be checked by additional experimental verifications.
We do not exclude this possibility, founding our viewpoint on the
papers~\cite{Hayashi,AVD,Kirch,EV-DVA}.

Finally, one
should note that we agree with the author of the cited
work~\cite{Ohanian}, see Eq. (4), about the gauge {\it non}-invariance of
the division of the angular momentum of the electromagnetic field into the
``orbital" and ``spin" part (\ref{spin}). We proved again
that for the antisymmetric tensor field ${\bf J} \sim \int d^3 {\bf x}\,
{\bf E}\times {\bf A}$. So, what people actually did (when spoken about
the Ogievetski\u{\i}-Polubarinov-Kalb-Ramond field is:
When $N=m$ they considered the gauge part of the 4-vector field functions.
Then, they equated ${\bf A}$ containing the transverse modes on choosing
$p_r =p_l =0$ (see formulas (\ref{vp12})).\footnote{The reader, of course,
can consider equating by the usual gauge transformation, $A^\mu
\rightarrow A^\mu +\partial^\mu \chi$.} Under this choice the ${\bf E}
({\bf p}, 0)$ and ${\bf B} ({\bf p}, 0)$ are equal to zero in massless
limit.  But, the gauge part of $u^\mu ({\bf p}, 0)$ is not. The spin
angular momentum can still be zero.
When $N=1$ the situation may be the same because of the different form
of dynamical equations and the Lagrangian. So, for those who prefer
simpler consideration it is enough to regard all possible states of
4-potentials/antisymmetric tensor field in the massless limit in the
calculation of physical observables. Of course, I would like to repeat, it
is not yet clear and it is not yet supported by reliable experiments
whether the third state of the 4-vector potential/antisymmetric tensor
field has physical significance and whether it is observable.

\section{Conclusions}

In conclusion, I calculated the Pauli-Lubanski vector of relativistic
spin on the basis of the N\"otherian symmetry
method~\cite{Corson,Barut,Bogoliubov}.  Let me recall that it is connected
with the angular momentum vector, which is conserved as a consequence of
the rotational invariance. After explicit~\cite{Hayashi} (or
implicit~\cite{AVD}) applications of the constraints (the generalized
Lorentz condition) in the Minkowski space, the antisymmetric tensor field
becomes `{\it longitudinal}' in the meaning that the angular momentum
operator is equated to zero (this interpretation was attached by the
authors of the works~\cite{Ogievet,Hayashi,AVD}).  I proposed
one of the possible ways to resolve this
apparent contradiction with the Correspondence
Principle in refs.~\cite{DVO1,DVO2,DVO3,DVO4} and in several unpublished
works\cite{DVO96}.
The present article continues and sums up this research.  The achieved
conclusion  is:  the antisymmetric tensor field can describe both the
Maxwellian $j=1$ field and the Ogievetski\u{\i}-Polubarinov-Kalb-Ramond
$j=0$ field.  Nevertheless, I still think that the physical nature of the
$E=0$ solution discovered in ref.~\cite{Gian}, its connections with the
so-called ${\bf B}^{(3)}$ field, ref.~\cite{Evans1,EV-DVA}, with
Avdeev-Chizhov $\delta^\prime$- type transversal solutions~[23b], which
cannot be interpreted as relativistic particles, as well as with my
concept of $\chi$ boundary functions, ref.~\cite{DVO4}, are not completely
explained until now.   Finally, while I do not have any intention of
doubting the theoretical results of the ordinary quantum electrodynamics,
I am sure that the questions put forth in this note (as well as in
previous papers of both mine and other groups) should be explained
properly.

\section{\bf Acknowledgements.}

As a matter of fact, the physical content of my series of papers has
been inspired by remarks of the referees of IJMPA (1994), FP and FPL
(1997) and HPA (1998).  I am thankful to Profs. A. E. Chubykalo, Y. S.
Kim, A.~F.~Pashkov and S.  Roy for stimulating discussions.  Several
papers of other authors which are devoted to a consideration of the
similar topics, but from very different standpoints, were motivations for
revising the preliminary versions of manuscripts.  I am
delighted by the referee reports on the papers~\cite{DVO1,DVO2,DVO3,DVO4}
from ``Journal of Physics A".  In fact, they helped me to learn many
useful things.

I am grateful to Zacatecas University for a professorship.
This work has been supported in part by the Mexican Sistema
Nacional de Investigadores and the Programa de Apoyo a la Carrera
Docente.

\nonumsection{References}

\end{document}